\documentclass[prl,superscriptaddress,showpacs,twocolumn]{revtex4}
\usepackage{epsfig}

\begin{document}
\title{Domain wall creep in epitaxial ferroelectric Pb(Zr$_{0.2}$Ti$_{0.8}$)O$_3$ thin films}
\author{T. Tybell}
\affiliation{DPMC, University of Geneva, 24 Quai E. Ansermet, 1211
Geneva 4, Switzerland}
\affiliation{Department of Physical
Electronics, Norwegian University of Science and Technology,
N-7491 Trondheim, Norway}
\author{P. Paruch}
\affiliation{DPMC, University of Geneva, 24 Quai E. Ansermet, 1211
Geneva 4, Switzerland}
\author{T. Giamarchi}
\affiliation{Laboratoire de Physique des Solides, CNRS-UMR 8502,
UPS B\^at. 510, 91405 Orsay France}
\author{J.-M. Triscone}
\affiliation{DPMC, University of Geneva, 24 Quai E. Ansermet, 1211
Geneva 4, Switzerland}
\date{\today}
\begin{abstract}
Ferroelectric switching and nanoscale domain dynamics were
investigated using atomic force microscopy on monocrystalline
Pb(Zr$_{0.2}$Ti$_{0.8}$)O$_3$ thin films. Measurements of domain
size versus writing time reveal a two-step domain growth
mechanism, in which initial nucleation is followed by radial
domain wall motion perpendicular to the polarization direction.
The electric field dependence of the domain wall velocity
demonstrates that domain wall motion in ferroelectric thin films
is a creep process, with the critical exponent $\mu$ close to 1.
The dimensionality of the films suggests that disorder is at the
origin of the observed creep behavior.
\end{abstract}
\maketitle

Understanding the propagation of elastic objects driven by an
external force in the presence of a pinning potential is a key to
the physics of a wide range of systems, either periodic, such as
the vortex lattice in type II superconductors
\cite{blatter_vortex_review}, charge density waves
\cite{gruner_revue_cdw} and Wigner crystals
\cite{andrei_wigner_2d}, or involving propagating interfaces, such
as growth phenomena \cite{kardar_review_lines}, fluid invasion
\cite{wilkinson_invasion} or magnetic domain walls
\cite{lemerle_domainwall_creep}. In particular, the response to a
small external force is of special theoretical and practical
interest. It was initially believed that thermal activation above
the pinning barriers should lead to a linear response at finite
temperature \cite{anderson_kim}. However, it was subsequently
realized that a pinning potential, either periodic
\cite{blatter_vortex_review} or disordered,
\cite{ioffe_creep,nattermann_rfield_rbond,blatter_vortex_review,chauve_creep_long},
can lead to diverging barriers and thus to a non-linear response,
nicknamed creep, where the velocity is of the form $v \propto
\text{exp}(-\beta R (f_c/f)^\mu)$. $\beta$ is the inverse
temperature, $R$ a characteristic energy and $f_c$ a critical
force. The dynamical exponent $\mu$ reflects the nature of the
system and of the pinning potential. Despite extensive studies of
the creep process in periodic vortex systems
\cite{blatter_vortex_review}, precise determination of the
exponents has proven difficult, given the many scales present in
this problem and the range of voltage needed to check the creep
law \cite{fuchs_creep_bglass}. For interfaces, a quantitative
check of the creep law has been done recently in ultrathin
magnetic films \cite{lemerle_domainwall_creep}, where an exponent
$\mu=0.25$ has been measured in very good agreement with the
expected theoretical value for this system. Quantitative studies
of this phenomena in other microscopic systems with other pinning
potentials are clearly needed.

\begin{figure}
\begin{center}
\includegraphics[scale=0.45]{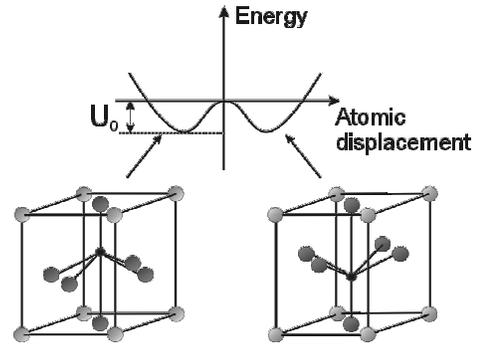}
\end{center}
\caption{ Schematic of a perovskite ferroelectric, characterized
by two oppositely polarized ground states, separated by an energy
barrier $U_0$. For Pb(Zr$_{0.2}$Ti$_{0.8}$)O$_3$, the corner Pb
ions and the center Ti/Zr ions are positively charged, and the
face O ions are negatively charged.} \label{fig1}
\end{figure}

In this respect, ferroelectric materials are of special interest.
These systems possess two symmetrically equivalent ground states
separated by an energy barrier $U_0$, as illustrated in
Fig.~\ref{fig1} for a tetragonal perovskite structure. Each state
is characterized by a stable remanent polarization, reversible
under an electric field.
Regions of different polarization are separated by elastic domain
walls. The application of an electric field favors one
polarization state over the other, by reducing the energy
necessary to create a nucleus with a polarization parallel to the
field, and thus promotes domain wall motion. In addition to pure
theoretical interest, understanding the basic mechanism of domain
wall motion in ferroelectrics has practical implications for
technological applications, such as high-density memories. In bulk
ferroelectrics, switching and domain growth were inferred to occur
by stochastic nucleation of new domains at the domain boundary, a
behavior observed in BaTiO$_3$ and triglycine sulphate by studies
using combined optical and etching techniques
\cite{merz_nonlinear_ferroelectrics,fatuzzo_nonlinear_ferroelectrics,%
miller_nucleation_ferroelectrics}. Early analysis of such motion
reported a field dependence of the domain wall speed, $v \sim
\exp{[-1/E]}$. This bulk system behavior was explained by assuming
that domain walls propagate via nucleation along 180$^{\circ}$
domain boundaries.

In this Letter, we report on studies of ferroelectric domain wall
motion in atomically flat single crystal
Pb(Zr$_{0.2}$Ti$_{0.8}$)O$_3$ films by atomic force microscopy
(AFM), allowing noninvasive investigation of domain dynamics with
nanometer resolution. In this model system, we identify domain
wall motion to be a disorder-controlled creep process. The
dynamical exponent $\mu$ is found to be close to 1. The activation
energy increases significantly from 0.5MV/cm to 1.3MV/cm as the
film thickness is reduced from 810\AA  to 290\AA.

The ferroelectric materials investigated were epitaxial c-axis
oriented Pb(Zr$_{0.2}$Ti$_{0.8}$)O$_3$ thin films, RF-magnetron
sputtered onto conducting (100) Nb-doped SrTiO$_3$ substrates
\cite{tybell_ferroelectrics_films,triscone_ferroelectrics_films}.
This system allows precise control of film thickness and
crystalline quality, exhibiting atomically smooth surfaces with a
polarization vector parallel or antiparallel to the c-axis
\cite{tybell_ferroelectrics_films,triscone_ferroelectrics_films}.
To study switching dynamics in these films, we used a conductive
AFM tip to artificially modify domain structure
\cite{tybell_ferroelectrics_films,guthner_afm_ferroelectrics,ahn_afm_ferroelectrics}.
Domains were polarized by applying a voltage pulse across the
ferroelectric film, between the tip and the substrate. Their sizes
were subsequently measured by piezoelectric microscopy
\cite{guthner_afm_ferroelectrics}, as a function of pulse width
and amplitude. For every pulse width, we used 12 V pulses to
polarize an array of 16 domains, and calculated an average domain
size based on their vertical and horizontal radii. The rms error
was $\sim$10\%. All domains studied were written in a uniformly
polarized area.
\begin{figure}
\begin{center}
\includegraphics[scale=0.6]{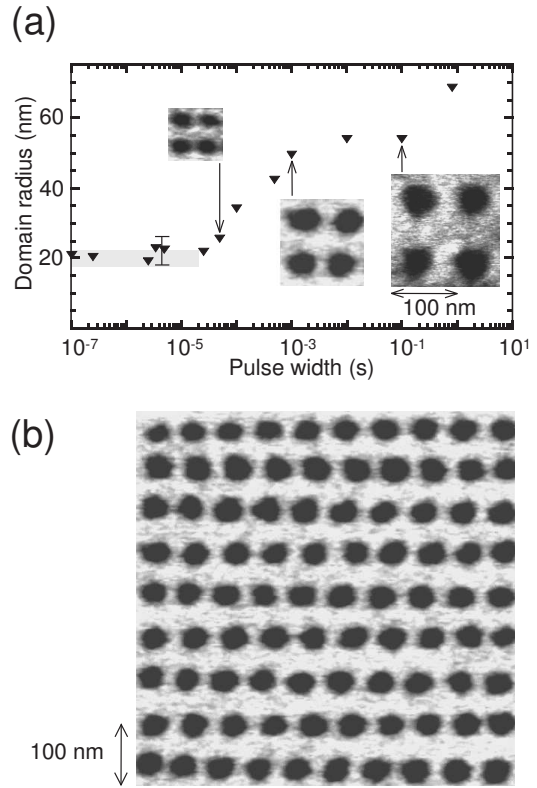}
\end{center}
\caption{a) Domain size increases logarithmically with pulse
widths longer than $\sim$20$\mu$s, and saturates for shorter times
as indicated by the shaded area. In all cases, the domains are
homogeneous and no random nucleation is detected, as shown by
three piezoelectric images of domain arrays on a 370{\AA} thick
film written with 50${\mu}$s, 1ms, and 100ms voltage pulses. The
scale is the same in all images.   In b), a large array shows that
domains are only centered where the AFM tip was positioned during
writing.} \label{fig2}
\end{figure}
Fig.~\ref{fig2}a shows the domain radius as a function of pulse
width, and three piezoelectric images of ferroelectric domain
arrays written with 50$\mu$s, 1ms, and 100ms voltage pulses. As
can be seen, varying the writing time (the pulse width) markedly
changes the size of the AFM-written domains. We observe that
domain radius increases logarithmically with increasing writing
time for times longer than $\sim$20$\mu$s. Below 20$\mu$s, and
down to 100ns, the shortest times investigated, domain radius is
found to be constant and approximately equal to 20nm, as shown by
the shaded area in Fig.~\ref{fig2}a
\cite{paruch_afm_ferroelectrics}. All of our data suggest that
this minimum domain size is related to the typical tip size used
for the experiments, whose nominal radius of curvature is
$\sim$20-50nm. In previous work, we have observed that domain size
also depends linearly on writing voltage, above a threshold
related to the coercive field \cite{paruch_afm_ferroelectrics}. A
detailed analysis of the data reveals only well-defined
homogeneous domains with regular spacing, as can be seen in the
piezoelectric image of a regular 90-domain array, written with 1ms
pulses in Fig.~\ref{fig2}b. We note that the topographic image of
the same area is featureless, with a rms roughness of $\sim$0.2nm.
Within our $\sim$5nm resolution we do not detect any randomly
nucleated domains. The data thus suggest a two-step domain
switching process in which nucleation, originating directly under
the AFM tip, is followed by radial motion of the domain wall
outwards, perpendicular to the direction of polarization.

\begin{figure}
\begin{center}
\includegraphics[scale=0.55]{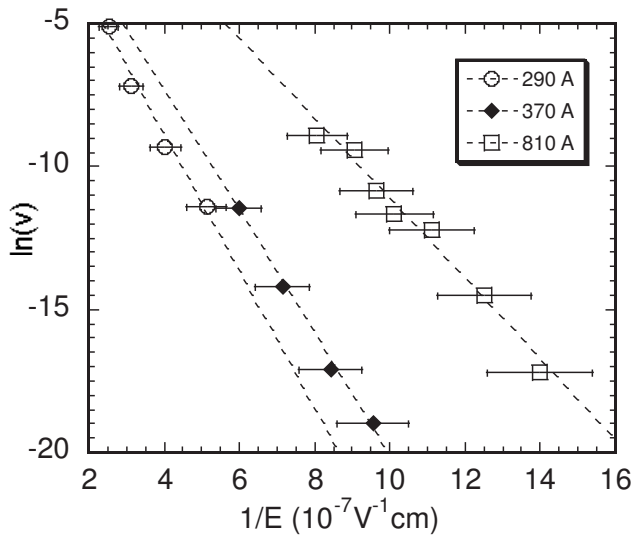}
\end{center}
\caption{Domain wall speed as a function of the inverse applied
electric field for 290{\AA}, 370{\AA}, and 810{\AA} thick samples.
The data fit well to $v \sim
\exp{[-\frac{R}{k_BT}\left(\frac{E_0}{E}\right)^\mu]}$ with $\mu =
1$, characteristic of a creep process. } \label{fig3}
\end{figure}

To analyze this lateral domain wall motion, we note that the force
exerted on the wall is given by the electric field $E$. To obtain
the electric field distribution, we model the tip as a sphere,
with a radius $a$. The potential on the ferroelectric surface at a
distance $r$ from the tip is then $\Phi \sim \frac{Va}{r}$, and
the local field across the ferroelectric $E = \frac{Va}{rd}$, $V$
being the applied tip bias and $d$ the film thickness. This
equation will allow us to relate the change in domain size to the
local electric field near the domain boundary. By writing arrays
with different pulse widths, and subsequently calculating the
average domain size for a given time, we can extract the speed of
the domain wall as $v = \frac{r(t_2)-r(t_1)}{t_2-t_1}$ and the
corresponding electric field $E(r)$ where $r = (r(t_1)+r(t_2))/2$.
Fig.~\ref{fig3} shows the wall speed as a function of the inverse
field for three different film thicknesses. The data fits well to
a creep formula
\begin{equation}
 v \sim \exp{-\frac{R}{k_BT}\left(\frac{E_0}{E}\right)^\mu}
\end{equation}
with $\mu = 1$. The exact dynamical exponent $\mu$ is found to be
1.12, 1.01 and 1.21 for the 290{\AA}, 370{\AA}, and 810{\AA} thick
films respectively,with an estimated 10{\%} rms error on the field
\footnote{The 10\% rms error in the local electric field is given
by the precision of the domain size measurements. We used 20nm as
the AFM tip radius of curvature.}. We find the effective
``activation energy'' $\big(R/(k_BT)\big)^{\frac{1}{\mu}}E_0$ to
be 1.321MV/cm, 1.305MV/cm, and 0.506MV/cm for 290{\AA}, 370{\AA},
and 810{\AA} thick films, typically 2 orders of magnitude larger
than the applied fields during the polarization process. We note
that during AFM writing, the effective field across the
ferroelectric is approximately 10 times smaller than the field E
\footnote{The exact magnitude of the electric field across the
ferroelectric is difficult to quantify, because of a possible gap
between the film surface and the tip
\cite{hidaka_afm_ferroelectrics}, and the unknown and gradually
changing tip shape. Local piezoelectric hysteresis measurements on
120\AA - 800\AA thick films show that the minimum switching field
is $\sim$16 - 6 times larger than the bulk coercive field. We
therefore correct the calculated field E by a factor of 10 to
estimate the effective field across the film in our spherical tip
model.}. This has no effect on the exponent $\mu$ governing the
exponential velocity dependence, but decreases the effective
``activation energy'' by a factor of 10, which has been taken into
account in the values given above.

Let us now consider the possible microscopic origins of the
observed creep behavior. Creep phenomena are a consequence of
competition between the elastic energy of a propagating interface,
tending to keep it flat, and a pinning potential, preventing it
from simply sliding when submitted to an external force. The
dynamical exponent in the creep scenario depends on both the
dimensionality of the system, and the nature of the pinning
potential. Although creep processes are generally associated with
the glassy behavior of disordered systems, they can also be
observed in a periodic potential if the dimensionality of the
object is larger than or equal to 2. For thick films, as for bulk
ferroelectrics, the domain wall is a two dimensional object. In
this case, from free-energy considerations, and neglecting the
anisotropic dipole field present in a ferroelectric, one would
expect the exponent to be $\mu=1$ \cite{blatter_vortex_review}. In
PbTiO$_3$, it has been shown theoretically that domain wall energy
depends upon whether the wall is centered on a Pb or Ti plane
\cite{poykko_periodic_potential_ferroelectrics,meyer_periodic_potential_ferroelectrics},
giving rise to an intrinsic periodic pinning potential. One
possible explanation could thus be that the observed creep is due
to the motion of the two-dimensional wall in this periodic
potential. Note that this scenario is a generalization of the
nucleation model developed for bulk ferroelectrics
\cite{miller_nucleation_ferroelectrics}. In order to test this
hypothesis we calculated the size of the critical nucleus, using
the formula derived by Miller and Weinreich
\cite{miller_nucleation_ferroelectrics}. To estimate $l^\ast$, the
critical length along the c-axis, we used the standard remanent
polarization, lattice parameters and dielectric constant values
for PZT , the 169mJ/m$^2$ domain wall energy derived for PbTiO$_3$
\cite{meyer_periodic_potential_ferroelectrics}, and the corrected
values for the electric field across the ferroelectric film.  We
find $l^\ast$ to vary, depending on the field range, between
200{\AA} and 500{\AA}, 600{\AA} and 1100{\AA}, and 900{\AA} and
1700{\AA} for the 290{\AA}, 370{\AA} and 810{\AA}. The critical
nucleus would thus need to be larger than the thickness of the
system. Furthermore, the effective ``activation energy''
calculated for the nucleation model is two orders of magnitude
greater than the 0.5-1.3MV/cm determined experimentally.
Calculations directly starting from the periodic potentials given
in \cite{meyer_periodic_potential_ferroelectrics} lead to similar
conclusions. These results strongly suggest that the films are in
a two dimensional limit, and that the nucleation model or
equivalently the motion through a periodic potential does not
adequately explain the experimental data.

The creep behavior in the film thus has to be due to disorder in
the system, and thereby to the glassy characteristics of a
randomly pinned domain wall, with the creep exponent dependent on
the nature of the disorder.  Defects locally modifying the
ferroelectric double well depth $U_0$ and giving rise to a
spatially varying pinning potential would lead to a ``random
bond'' scenario similar to the one for the ferromagnetic domain
walls \cite{lemerle_domainwall_creep}. The exponent $\mu$ would be
$\mu = \frac{d- 2 + 2\zeta}{2-\zeta}$ where $\zeta$ is a
characteristic wandering exponent and $d$ the dimensionality of
the wall. For one dimensional domain walls $\mu =1/4$ whereas for
two dimensional ones $\mu\sim 0.5-0.6$, hardly compatible with the
data. If, however, the defects induce a local field, asymmetrizing
the double well, or if there are spatial inhomogeneities in the
electric field, the situation is different. For such a ``random
field'' scenario $\zeta = \frac{4-d}{3}$
\cite{fisher_functional_rg} and thus leads to $\mu = 1$ for
$1<d<4$. This scenario is therefore compatible with the observed
data. However, further study determining the wandering exponent
would be needed to ascertain this precise point \footnote{Only
short range elastic forces are considered in the above discussion.
Long range interactions would change the elastic energy, and
therefore the dynamics of the system, and hence the exponent
$\mu$. In this case, the random bond scenario could also lead to
different exponents and possibly to $\mu$ close to 1.}.

\begin{figure}
\begin{center}
\includegraphics[scale=0.6]{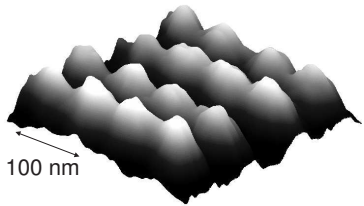}
\end{center}
\caption{Piezoelectrically imaged high density ferroelectric
domain array at 28Gbit/cm$^2$, written with 500ns pulses. }
\label{fig4}
\end{figure}

Finally, we note the applications of these results to
technological developments. The fact that the domain walls exhibit
creep motion with a relatively large exponent $\mu=1$ implies a
strong stability of ferroelectric domains in thin films, since the
induced speed of the domain wall becomes exponentially small as
the driving force goes to zero. All domains studied in this work
were stable under ambient conditions for the entire 7 day duration
of the experiment. Furthermore, sub-$\mu$m wide line-shaped
domains were stable up to one month
\cite{marre_stability_walls_ferroelectrics}. Previously, large,
regular arrays of ferroelectric domains with densities of
$\sim$6Gbit/cm$^2$ have been reversibly written
\cite{paruch_afm_ferroelectrics}, and densities up to
150Gbit/cm$^2$ have been extrapolated from sizes of individual
domains \cite{maruyama_arrays_ferroelectrics}. This work
identifies the key parameters controlling domain size: the
strength, duration, and confinement of the applied electric field.
These can be exploited to increase the information storage density
in ferroelectric arrays. By using short voltage pulses, for which
the resulting domain size is independent of the writing time, and
domain-domain separation as small as 10nm, regular arrays with
densities of the order $\sim$30Gbit/cm$^2$ could be written.
Fig.~\ref{fig4} shows such an array written on a 370{\AA} thick
sample with a density of 28Gbit/cm$^2$. Furthermore, it has been
suggested \cite{woo_stability_ferroelectrics} that the minimum
stable domain size is given by the film thickness. Our data,
however, show that the minimum stable domain size, $\sim$40nm, is
independent of the thickness used in this study; rather, it is
given by the area over which the electric field is applied during
the polarization process. Therefore, by confining the field to a
smaller area, domain wall creep will be limited, leading to
smaller domains and hence even higher information densities.

In conclusion, our studies demonstrate that lateral domain wall
motion in ferroelectric thin films is a creep process, governed by
a characteristic dynamical exponent $\mu$ close to $1$.   The
dimensionality of our films suggests that disorder is at the
origin of the observed creep behavior, which inherently explains
the measured $\text{exp}(-1/E)$ dependence of the domain wall
speed. The activation energy is found to be around 1MV/cm,
decreasing with increasing film thickness. Finally, these results
suggest a high degree of stability for ferroelectric domains in
low electric fields and identify the key parameters controlling
domain size.

\begin{acknowledgments}
The authors would like to thank Gianni Blatter, Dima Geshkenbein,
and Piero Martinoli for enriching discussions. Special thanks to
Steve Brown for carefully reading the manuscript, and to Daniel
Chablaix for useful technical developments. This work was
supported by the Swiss National Science Foundation through the
National Center of Competence in Research ``Materials with Novel
Electronic Properties-MaNEP'' and Division II.
\end{acknowledgments}

\bibliographystyle{prsty}

\end{document}